\documentclass[twocolumn,showpacs,amsmath,amssymb,superscriptaddress]{revtex4}
\usepackage{amsmath,amssymb,color,latexsym,natbib,graphicx}
\usepackage{ulem,color,soul}
\usepackage{dcolumn}
\usepackage{bm}% bold math
\def\ie{{\it i.e.}\ }

\def\b0{{\bf b_0}}
\def\kpe{{k_\perp}}
\def\kpa{{k_\parallel}}

\def\ep{{\bf \hat{e}}_{\parallel}}

\def\be{\begin{equation}}
\def\ee{\end{equation}}
\def\ba{\begin{eqnarray}}
\def\ea{\end{eqnarray}}

\def \pmbtext#1{\leavevmode
     \setbox0\hbox{#1}
     \kern0,4pt \copy0 \kern-\wd0
     \kern-0,2pt \raise0,3pt \box0 }
     
\begin{document}

\preprint{1}

\title{Direct Evidence of the Transition from Weak to Strong MHD Turbulence}
\author{Romain Meyrand}
\affiliation{Space Sciences Laboratory, University of California, Berkeley, CA 94720, USA}
\affiliation{LPP, \'Ecole Polytechnique, F-91128 Palaiseau Cedex, France}
\email{romain.meyrand@ssl.berkeley.edu}
\author{S\'ebastien Galtier}
\affiliation{LPP, \'Ecole Polytechnique, F-91128 Palaiseau Cedex, France}
\affiliation{Univ. Paris-Sud, Orsay, France}
\email{sebastien.galtier@lpp.polytechnique.fr}
\author{Khurom H. Kiyani}
\affiliation{LPP, \'Ecole Polytechnique, F-91128 Palaiseau Cedex, France}
\affiliation{Centre for Fusion, Space and Astrophysics, University of Warwick, Coventry CV4 7AL, UK}
\email{khurom.kiyani@lpp.polytechnique.fr}

\date{\today}
%%%%%%%%%%%%%%%
\begin{abstract}
One of the most important predictions in magnetohydrodynamics (MHD) is that in the presence of a uniform magnetic field $b_0 \ep$ a transition from weak to strong wave turbulence should occur when going from large to small perpendicular scales. This transition is believed to be a universal property of several anisotropic turbulent systems. We present for the first time direct evidence of such a transition using a decaying three-dimensional direct numerical simulation of incompressible balanced MHD turbulence with a grid resolution of $3072^2 \times 256$. From large to small-scales, the change of regime is characterized by i) a change of slope in the energy spectrum going from approximately $-2$ to $-3/2$; ii) an increase of the ratio between the wave and nonlinear times, with a critical ratio of $\chi_{c}\sim 1/3$; iii) a modification of the iso-contours of energy revealing a transition from a purely perpendicular cascade to a cascade compatible with the critical balance type phenomenology, and iv) an absence followed by a dramatic increase of the communication between Alfv\'en modes. The changes happen at approximately the same transition scale and can be seen as manifest signatures of the transition from weak to strong wave turbulence. 
Furthermore, we observe a significant non-local three-wave coupling between strongly and weakly nonlinear modes resulting in an inverse transfer of energy from small to large-scales.
\end{abstract}
%%%%%%%%%%%%%%%
\maketitle

%%%%%%%%%%%%%%%
\paragraph*{Introduction.} 
Waves are ubiquitous in natural systems. Although waves are a basic phenomenon well understood for years, the nonlinear behavior 
of a large ensemble of waves is still the subject of intense research \cite{Kolmakov,Falcon,Mininni11,Meyrand13,mordant,campagne15}. When a turbulent state is developed in the presence of waves one may distinguish two regimes: weak wave turbulence and strong wave turbulence. In the first case, the regime can be described analytically by a classical technique based on perturbative developments \cite{naza}. Exact solutions of the resulting kinetic equations, corresponding to power law spectra, can then also be derived \cite{Zakh65,BN67,Dyachenko,Galtier03,lvov03,G2014}. In the second case, we mainly have phenomenological models \cite{I64,K65,zhou}; among which we find the critical balance (CB). Since its inception, originally in MHD \cite{H84,GS95}, CB has become a popular model in astrophysics and, is believed to be also applicable to other systems such as electron MHD, rotating hydrodynamics and stratified flows \cite{CL04,GPM05,NS11}. In incompressible MHD, CB supposes the existence of a mean magnetic guide field $\b0$ (which will be normalized to velocity units hereafter) along which propagate Alfv\'en waves in both directions parallel ($\parallel$) to $\b0$. Both linear and nonlinear physics are affected by $\b0$ with the development of a high degree of anisotropy such that energy will mainly transfer, or cascade, in the perpendicular ($\perp$) direction to $\b0$ \cite{shebalin83,Grappin10}. In such a situation, the following inequality is satisfied $\kpe \gg \kpa$. As a result of this strong anisotropy, the (local) nonlinear time-scale becomes $\tau_{nl} \sim 1 / (\kpe b)$, whereas the linear Alfv\'en wave time-scale is $\tau_w \sim 1 / (\kpa b_0)$ (for the derivation, see the comments on Eqs. (\ref{mhd1})). 
The latter time-scale can be interpreted as the duration of a collision between two wave packets traveling in the opposite direction at the Alfv\'en speed $b_0$. The characteristic transfer time of energy $\tau_{tr}$ can, as far as dimensional analysis is concerned, be an arbitrary function of these two times -- an additional physical assumption is therefore necessary to fix the scaling. This additional assumption is furnished by the CB conjecture which assumes that at all scales in the inertial range $\tau_{nl} \sim \tau_w$. This physically means that an Alfv\'en wave packet suffers a deformation of the order of the wave packet itself in one collision. Two properties can be derived immediately from this assumption: (i) the axisymmetric energy spectrum is simply of the Kolmogorov type, \ie $b^2 / \kpe \sim \kpe^{-5/3}$, 
because the transfer time identifies to $\tau_{nl}$, 
and (ii) a non-trivial relationship exists between the parallel and perpendicular wave numbers in the form of 
$\kpa b_0 \sim \kpe b \sim \kpe^{2/3}$ \cite{H84,GS95}; in particular, this latter identity physically implies that anisotropy will increase at small-scales until the dissipation becomes dominant. The CB regime is drastically different from the weak wave turbulence one where, in the latter, many stochastic collisions are necessary to modify a wave packet significantly. In the case of weak turbulence, we have the inequality $\tau_w \ll \tau_{nl}$ and the transfer time becomes $\tau_{tr} \sim \tau_{nl}^2 / \tau_w$ \cite{I64,K65}. This transfer time can be interpreted as the time that it takes the cumulative perturbation (assumed to accumulate as a random walk) to become comparable to the amplitude of the wave packet itself. The resulting power law spectrum in MHD corresponds to $\kpe^{-2}$. This result, presented first in a non-rigorous phenomenological way \cite{NG97}, was then derived rigorously by a perturbative theory \cite{galtier00,galtier02}. 
In particular, it is shown that no transfer (cascade) is expected along the parallel direction, a result stemming from the three-wave resonance condition \cite{shebalin83}.
The necessary condition for the existence of weak MHD turbulence is that the time ratio
\be
\chi(\kpe,\kpa) \equiv {\tau_w \over \tau_{nl}} = {\kpe b \over \kpa b_0} \, 
\label{chi}
\ee
is small ($\ll 1$), whereas in CB it is of the order of one ($\sim 1$). If we substitute the weak turbulence spectrum $b^{2}/\kpe \sim \kpe^{-2}$ into 
Eq. (\ref{chi}), we see that $\chi$ is an increasing function of $\kpe$. Therefore, there exists a critical scale beyond which 
the weak turbulence cascade drives itself into a state which no longer satisfies the premise on which the theory is based. The dynamical breakdown of 
the weak turbulence description is expected to be followed by a saturation around one of $\chi$ because of the causal impossibility to 
maintain $\tau_w \gg \tau_{nl}$ \citep{schekochihin}. 
This means that for a sufficiently extended inertial range we should observe the transition from the weak turbulence regime to the CB one 
\citep{galtier00,verdini12,schekochihin}. Note, however, that CB may be refined by introducing the local dynamic alignment of the velocity and 
magnetic field fluctuations which corresponds to a modification of the nonlinear time-scale \cite{B06}. In this case the power law energy 
spectrum is expected to be $\sim\kpe^{-3/2}$. 

In this Letter, we present for the first time direct evidence of such a weak to strong transition, by means of a high resolution three-dimensional 
direct numerical simulation.

%%%%%%%%%%%%%%%
\paragraph*{Simulation setup.} 
The incompressible MHD equations, for our simulations, in the presence of a constant ${\bf b_0}$ are: 
\be
\partial_t {\bf z^\pm} \mp b_0 \partial_{\parallel} {\bf z^\pm} + {\bf z^\mp} \cdot \nabla \, {\bf z^\pm} = 
- {\bf \nabla} P_* + \nu_3 \Delta^3 {\bf z^\pm} \, , \, \, \label{mhd1}
\ee
where $\nabla \cdot {\bf z^\pm} = 0$, 
${\bf z^\pm}={\bf v} \pm {\bf b}$ are the fluctuating Els\"asser fields, ${\bf v}$ the plasma flow velocity, ${\bf b}$ the normalized magnetic field 
(${\bf b} \to \sqrt{\mu_0 \rho_0} \, {\bf b}$, with $\rho_0$ a constant density and $\mu_0$ the magnetic permeability), $P_*$ the total (magnetic 
plus kinetic) pressure, and $\nu_3$ a hyper-viscosity (a unit magnetic Prandtl number is taken). 
We see that the times $\tau_w$ and $\tau_{nl}$ are obtained respectively from the linear dispersive term and the nonlinear term on 
the LHS of Eqs. (\ref{mhd1}), assuming a balance ($z^+ \sim z^- \sim u \sim b$) and anisotropic ($\kpe \gg \kpa$) turbulence.

Eqs. (\ref{mhd1}) are computed using a pseudo-spectral solver called TURBO \citep{teaca,Meyrand12} with periodic boundary 
conditions in all three directions and with $3072^2 \times 256$ collocation points (the lower resolution being in the ${\bf b_0}$ 
direction where the cascade is reduced; however, the numerical box is not elongated and has an aspect ratio of one). The nonlinear 
terms are partially de-aliased using a phase-shift method. The initial state consists of isotropic magnetic and velocity field fluctuations 
with random phases such that the total cross-helicity, as well as the total magnetic and kinetic helicities, is zero (balanced and non-
helical turbulence). 
The kinetic and magnetic energies are equal to $1/2$ and localized at the largest scales of the system (wave numbers $k\in [2,4]$ are initially excited). We opt for a decaying turbulence mainly to avoid any artefact due to the external forcing \citep{biskamp2000}. Our analysis is systematically made at a time $t_{*}$ when the mean dissipation rate reaches its maximum, for which the turbulence is fully developed and the spectrum the most extended. Note that for $t > t_{*}$ the spectrum experiences a smooth self-similar decay which leads to a slow drift of the critical (transition) scale toward higher $k_{\perp}$ as the $\chi$ parameter is proportional to the amplitude of $\textbf{b}$.
We fix $\nu_3 = 4\times10^{-17}$ and $b_0=20$. Note that initially the energy of the 2D modes are taken to be zero in order to favor dynamics dominated by wave modes. With our (isotropic) initial conditions anisotropy will develop such that energy will fill the Fourier space with $\kpe \gg \kpa$. 

%%%%%%%%%%%%%%%
\paragraph*{Results.} 
\begin{figure}
\includegraphics[width=1.0\linewidth]{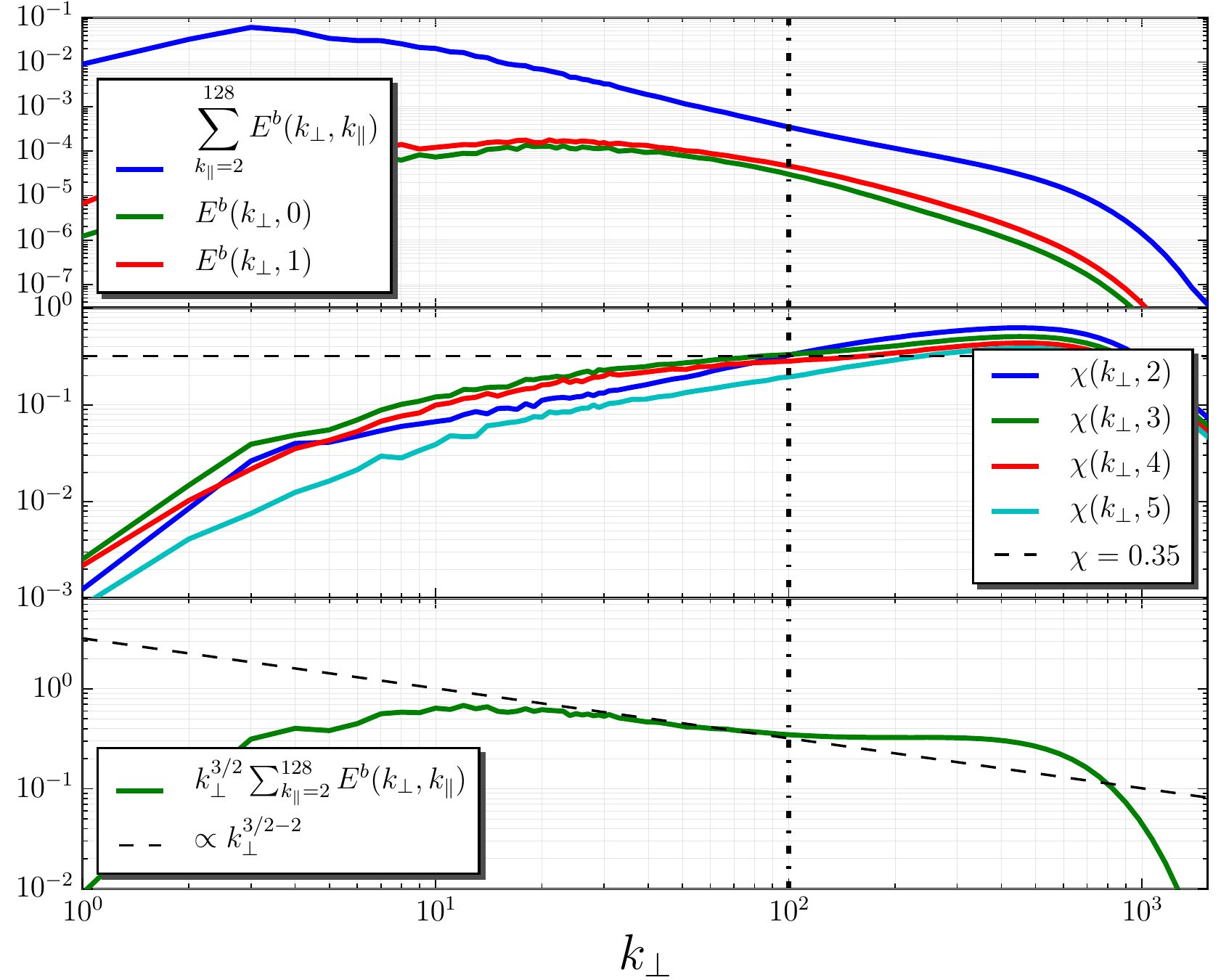}
\caption{Top: axisymmetric spectra of the magnetic energy at a given $\kpa=\{0,1\}$ and integrated over $\kpa$ (from $2$ to 
$128$). Middle: time ratio $\chi$ at a given $\kpa=\{2,3,4,5\}$. Bottom: integrated magnetic energy spectrum compensated by $\kpe^{3/2}$. 
The dashed line corresponds to a compensated spectrum with $\kpe^{-2}$. The vertical line marks the critical scale at which the transition is observed.}
\label{Fig1}
\end{figure}
We introduce the axisymmetric bi-dimensional magnetic energy spectrum $E^b(\kpe,\kpa)$ which is linked to the magnetic energy of the system ${\cal E}^b$ through the relation ${\cal E}^b = \iint E^b(\kpe,\kpa) d\kpe d\kpa$. It is well-known that in weak MHD turbulence the 2D mode ($\kpa=0$) has a singular role since it drives the turbulence although it is not a wave (see eg. \citep{Meyrand2015}). To make the distinction between the contributions of the 2D mode and the waves we have considered the spectrum $E^b(\kpe,\kpa)$ integrated from $\kpa=2$ to $128$ (the first plane $\kpa=1$ is found to be strongly coupled with the 2D mode like for enslaved turbulence \cite{Naza07}). The result is shown in Fig. \ref{Fig1} (top). At large-scale ($10<\kpe<100$) a spectrum compatible with weak turbulence is observed ($\sim \kpe^{-2}$). We then see that a transition seems to occur at a scale $\kpe \sim 100$ beyond which the spectrum becomes less steep. This transition happens considerably before the dissipative range which appears at $\kpe \sim 600$. The plots of the spectra for $\kpa=0$ and $1$ are also provided to show that there are similar, and significantly weaker in amplitude than the integrated spectrum. 
The time ratio (\ref{chi}) is shown in the middle panel of Fig. \ref{Fig1} for different (small) values of $\kpa$. For this evaluation of $\chi$, $b$ is given 
by $b = \sqrt{2 \kpa \kpe E^b(\kpe,\kpa)}$. In all cases, we see that $\chi(\kpe,\kpa) \ll 1$ at the largest scales as expected for the weak turbulence 
regime. The comparison with the spectra described above shows that a transition occurs when $\chi(\kpe,\kpa)$ approaches unity ($> 0.1$). For 
$\kpa>4$ we find that the higher $\kpa$, the smaller $\chi$, as expected from Eq. (\ref{chi}). 
Note that we do not observe at small-scales an extended plateau where $\chi \sim 1$ 
which could be explained by the lack of resolution, the fact that we did not consider the local mean magnetic field and/or the absence of dynamic alignment in the definition of the nonlinear time-scale \citep{Mallet15}. 
The last plot (bottom panel) shows the integrated spectrum compensated by $\kpe^{3/2}$. The transition is visible at $\kpe \sim 100$ with a change in slope going from approximately $\kpe^{-2}$ to $\kpe^{-3/2}$; this happens at $\chi_c \sim 1/3$.
\begin{figure}
\includegraphics[width=1.0\linewidth]{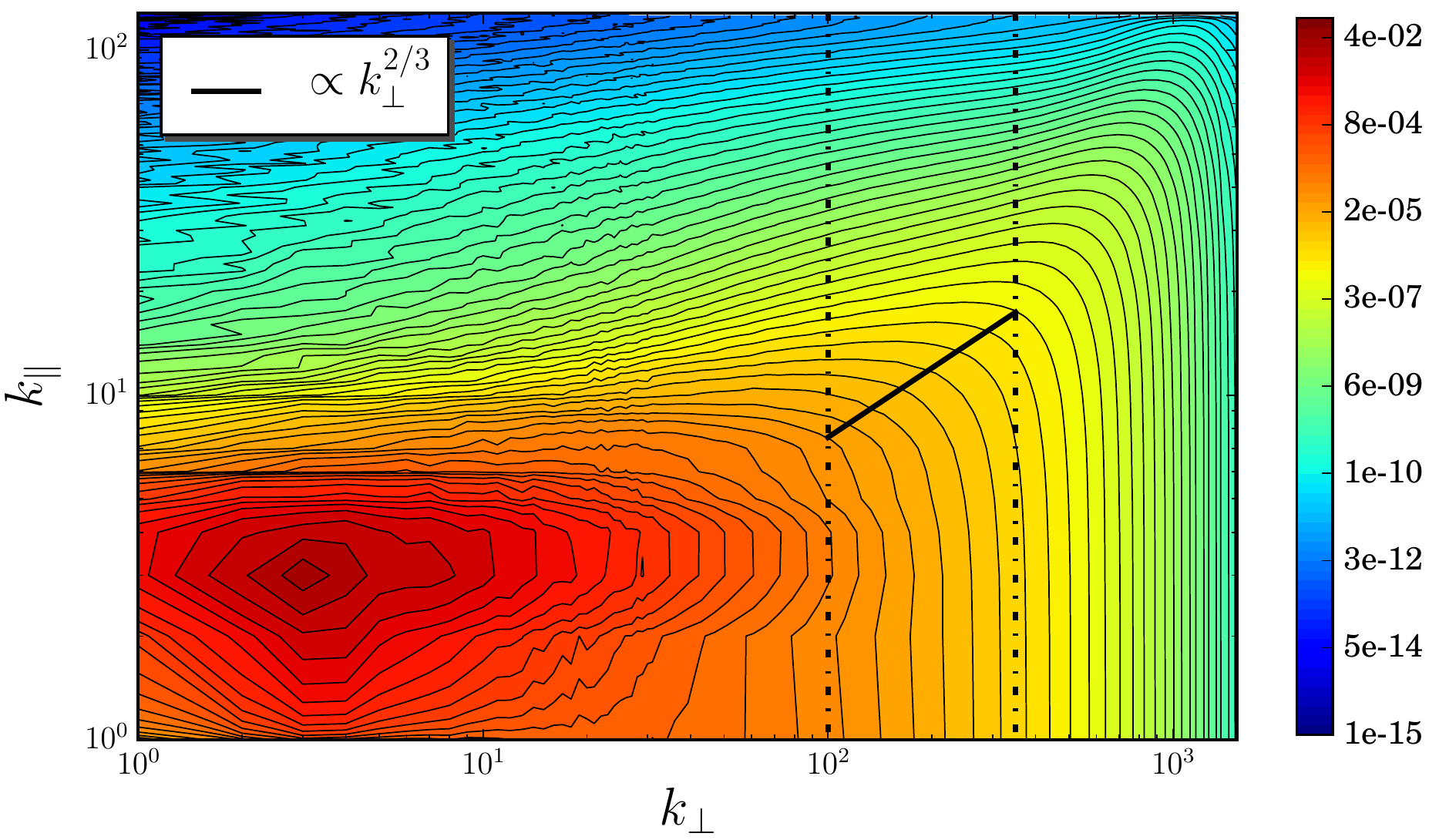}
\caption{Iso-contours (in logarithmic scale) of the bi-dimensional magnetic energy spectrum $E^b(\kpe,\kpa,t_*)$. 
A power law $\kpe^{2/3}$ is plotted for comparison in the region corresponding to strong wave turbulence (see Fig. \ref{Fig1}).}
\label{Fig2}
\end{figure}
Fig. \ref{Fig2} displays the iso-contours of the bi-dimensional magnetic energy spectrum. At large-scale the iso-contours are strongly elongated in the $\kpe$ direction meaning that the cascade is strongly anisotropic. At the transition scales ($\kpe \sim 100$) a drastic modification appears with an increase of the parallel transfer  and therefore a stretching of the iso-contours in the $\kpa$ direction. Interestingly, we may find a domain where the edge follows approximately a power law in $\kpe^{2/3}$. This means that the energy is mainly transferred along an oblique direction which corresponds to CB. When the dissipative scales are reached the stretching of the iso-contours in the $\kpa$ direction increases further  showing a propensity toward isotropization. 

\begin{figure}
\includegraphics[width=1.0\linewidth]{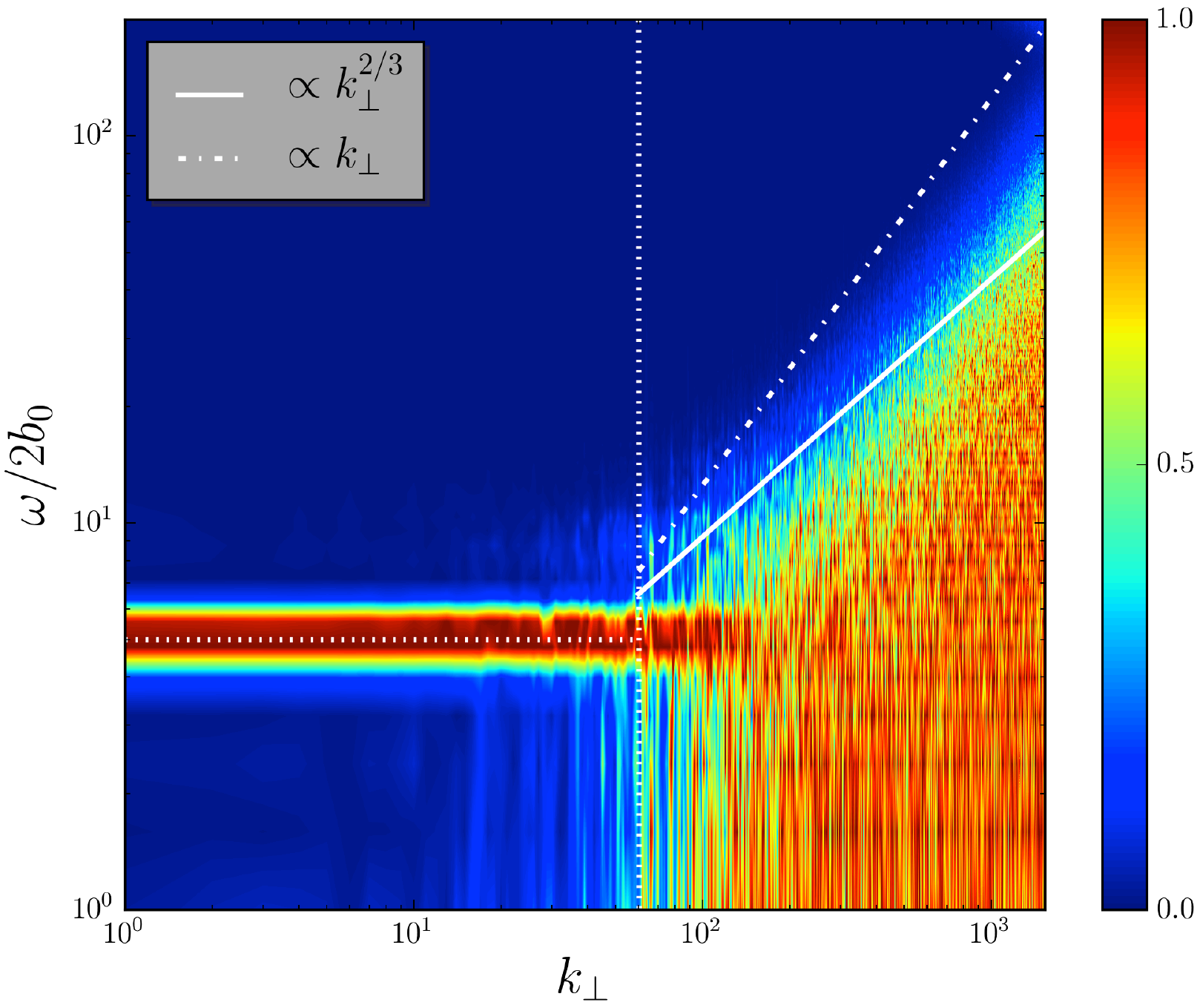}
\caption{Wavenumber--frequency spectrum of the magnetic energy $E^b(\kpe,\kpa=5,\omega)$. 
The color map is normalized to the maximal value of the spectrum at each fixed $\kpe$. 
%Note the presence of a factor $1/2$ in the frequency normalization because the energy (square of a field) is used. 
For comparison we plot $\kpe^{2/3}$ (solid) and $\kpe$ (dash-dot). 
The vertical dotted line marks the critical scale at which the transition is observed 
and the horizontal dotted line corresponds to $\omega/(2b_0)=5$}.
\label{Fig3}
\end{figure}

Although Fig.\ref{Fig1} and \ref{Fig2} may provide a first evidence of a transition from weak to strong turbulence, we want to find other signatures. 
The spectrogram (wavenumber--frequency spectrum) of the magnetic energy provides this additional information. To build such a spectrogram one 
follows, in Fourier space and over a window of time around $t_*$, the quantity $E^b(\kpe,\kpa,t) = \vert \hat b_x \vert^2 + \vert \hat b_y \vert^2 + 
\vert \hat b_z \vert^2 $, at a given $\kpa$ ($\kpa=5$) and a given $\kpe$ (from $1$ to $1536$). We then perform a time-Fourier transform of these $1536$ signals multiplied by a Hamming window and obtain $E^b(\kpe,\kpa=5,\omega)$. The result is shown in Fig. \ref{Fig3} (the kinetic energy is not shown but behaves similarly). As we can see, at large-scales ($\kpe < 60$) the signal is concentrated on a thick band localized around 
$\omega/(2b_0)=5$; thus the variable $\omega$ is closely related to $\kpa$ like the Alfv\'en wave dispersion relation $\omega_A = \kpa b_0$
(the factor $1/2$ in the frequency normalization is due to the use of the energy, a square of a field). 
That means the mode $\kpa=5$ communicates only with modes directly contiguous to it, \ie $\kpa=4,6$ and that 
the system is not able to redistribute the energy to a wide range of frequencies, a situation expected when the resonant triadic interactions dominate \cite{naza}. The cascade is then strongly anisotropic with a transfer mostly in the perpendicular direction \citep{Meyrand2015}.
The thickness of the band can be interpreted as nonlinear broadening due to the weak nonlinearity effects and shows that the resolution in the $\parallel$ direction is high enough to not fall into the discrete regime \cite{naza,nazaono}. 
From $\kpe \sim 60$ a drastic change appears: suddenly the energy spreads over a wide range of frequencies. This is the most spectacular evidence of the emergence of a strong wave turbulence regime in which the $\parallel$ cascade is not frozen and where 
the $\kpa=5$ mode becomes dynamically connected to a growing number of other Alfv\'en modes (\ie $\kpa \neq 5$) when one goes to higher $\kpe$. Besides this important property, we note in passing that the boundary which delimits the region where modes are dynamically connected follows a power law close to $\kpe^{2/3}$ which could be interpreted as a signature of CB since the balance condition implies $\omega \sim 1/ \tau_{nl}$. 
The plot shows, however, that frequencies are also excited below this boundary which would correspond to $\tau_w > \tau_{nl}$ 
(like in the solar wind \cite{Matthaeus2014}, but in apparent contradiction with a previous claim \citep{schekochihin} based on a heuristic description
and the assumption of local interactions).
At this stage, it is important to remind that to define the nonlinear time-scale we have implicitly assumed the locality of the interactions. 
Then, the previous observation could also be interpreted as the signature of non-local interactions.
Note finally that in the dissipative range ($\kpe > 600$) the boundary discussed above seems to follow a power law close to $\kpe$ which could be the signature of an isotropisation. 

\begin{figure}
\includegraphics[width=1.0\linewidth]{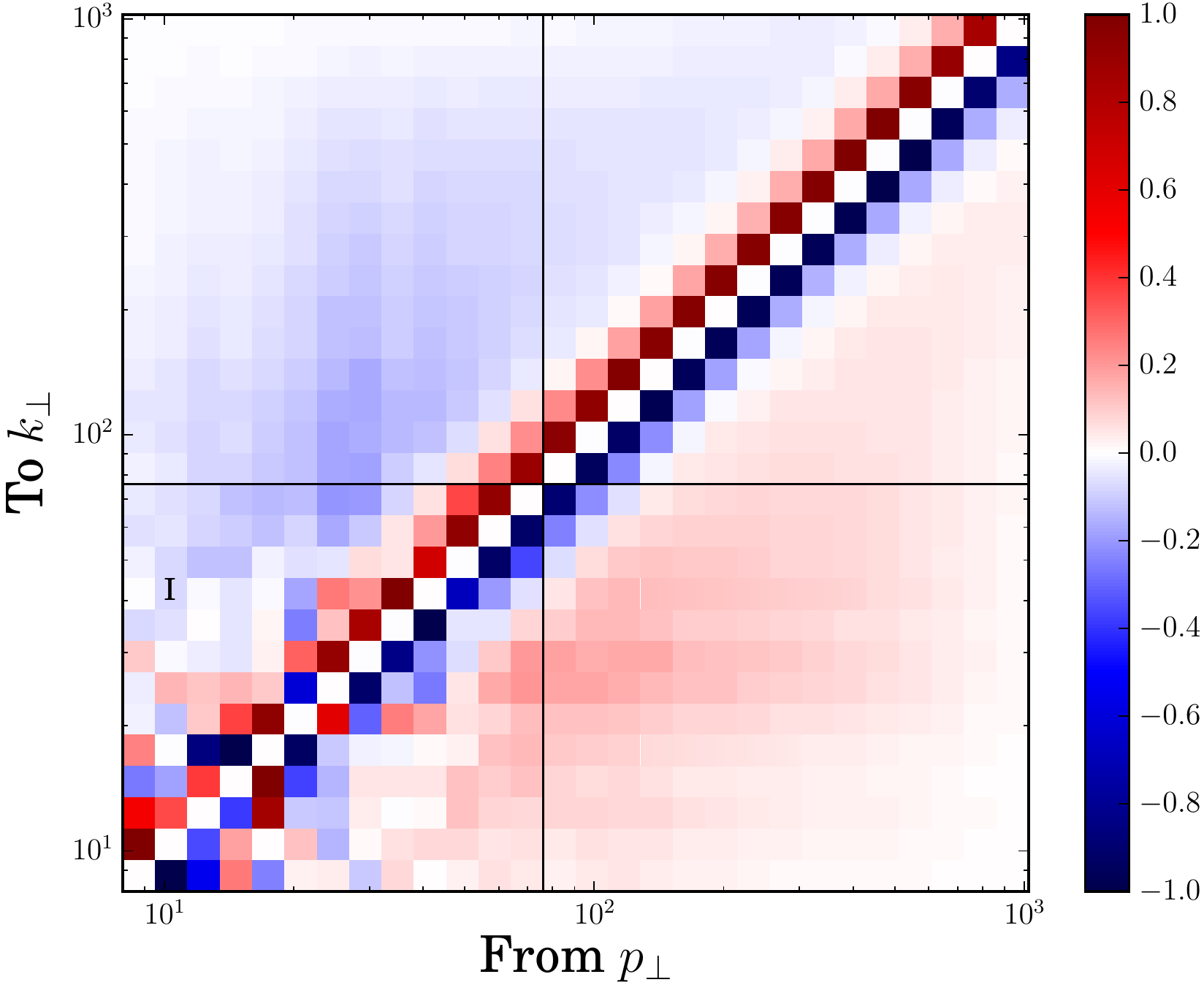}
\caption{Transfer functions $T(k,p)$ of the total energy normalized to the maximum absolute value of each wavenumber scale. 
The vertical and horizontal lines mark the critical scale at which the transition is observed (see Fig. \ref{Fig3}).}
\label{Fig4}
\end{figure}
The degree of locality of the perpendicular cascade can be investigated with the shell-to-shell energy transfer functions defined by: 
\begin{eqnarray}
\partial_t E^{u}(\textbf{k}) = \sum_{p}\left[T^{u}_{uu}(\textbf{k},\textbf{p})- T^{u}_{bb}(\textbf{k},\textbf{p})\right]-2\nu_{3} k^{6}E^{u}(\textbf{k}),\\
\partial_t E^{b}(\textbf{k}) =\sum_{p}\left[T^{b}_{bu}(\textbf{k},\textbf{p})-T^{b}_{ub}(\textbf{k},\textbf{p})\right]-2\nu_{3} k^{6}E^{b}(\textbf{k}),
\end{eqnarray}
where \citep{Debliquy2005,Carati2006,Alexakis2007}
\begin{equation}
T^{X}_{YZ}(\textbf{k},\textbf{p}) = \sum_{\textbf{q}}\mathfrak{Im}\lbrace[\textbf{k}\cdot \hat{\textbf{Z}}(\textbf{p}) ][\hat{\textbf{Y}}(\textbf{q})\cdot \hat{\textbf{X}}^{*}(\textbf{k})]\rbrace \delta_{\textbf{q}+\textbf{p},\textbf{k}},
\label{transfers}
\end{equation}
is the transfer function to the mode $\textbf{k}$ of field $\textbf{X}$ from mode $\textbf{p}$ of field $\textbf{Z}$, mediated by all possible triadic interactions with modes $\textbf{q}$ of fields $\textbf{Y}$ that respects the condition $\textbf{k}=\textbf{p}+\textbf{q}$. $\mathfrak{Im}$ denotes the imaginary part and $*$ the complex conjugate. 
To study the perpendicular cascade, we consider concentric cylindrical shells along $\b0$ with constant width on a logarithmic scale which we define as the region $k_{0}2^{n/4}\leq k_{\perp} \leq k_{0}2^{(n+1)/4}$ for the shells numbered $4\leq n\leq N$, where we set $k_{0}=4$ and $N=31$. 
The sum of transfer functions of the total energy ($T=T^{u}_{uu}-T^{u}_{bb}+T^{b}_{bu}-T^{b}_{ub}$) is displayed in Fig. \ref{Fig4}. While direct and 
local energy transfers dominate with transfers mainly concentrated around the diagonal $k_{\perp} = p_{\perp}$ (a result found in previous decaying 
MHD turbulence studies \citep{Debliquy2005,Alexakis2007}), one observes some inverse and non-local contributions connecting weak and strong 
modes. This behavior is revealed by the presence of negative and positive energy transfer respectively in the top-left and bottom right part of Fig. 
\ref{Fig4}. This result is new and important for the theory of MHD turbulence where this type of interaction has never been considered in the past.

%%%%%%%%%%%%%%%
\paragraph*{Discussion.} 
The transition from weak to strong wave turbulence when passing from large to small-scales is believed to be a universal property of several anisotropic turbulent systems with different underlying physics \citep{NS11,Yokoyama}.  To our knowledge -- and despite its importance -- this phenomenon has so far never been observed in direct numerical simulations of MHD. This has left crucial questions unanswered. As a result of the simulation conducted here, we provide in this Letter a direct validation of this cornerstone of anisotropic turbulence theory in the MHD case, and are now able to provide answers to some of these fundamental questions. 
It appears sufficient that the parameter $\chi(\kpe,\kpa)$ crosses the critical value $\sim 1/3$ for a given $\kpa$ plane to contaminate rapidly the others whatever their respective degree of nonlinearity. The spectral index and anisotropy of the total energy after the breakdown of the weak turbulence description is consistent with the establishment of CB. 
In addition, our results indicate that the transition involves blending and interaction between weakly {\it and} strongly nonlinear modes. This unexpected behavior revealed by the presence of non-local and inverse energy transfers suggests that the transition is not simply the juxtaposition of weak and strong wave turbulence as was thought until now. 
This result is potentially important for other systems where a transition form weak to strong wave turbulence is expected \cite{naza}.
%%%%%%%%%%%%%%%
\paragraph*{Acknowledgements.} 
The research leading to these results has received funding from the European Commission's 7th Framework Program (FP7/2007-2013) under the grant agreement SHOCK (project number 284515), from the ANR contract 10-JCJC-0403 and ANR-JC project THESOW. The computing resources were made available through the UKMHD Consortium facilities funded by STFC grant number ST/H008810/1. R.M  acknowledges the financial support from EU-funded Marie Curie-Sk\l{}odowska Global Fellowship.
\bibliographystyle{apsrev4-1}
\bibliography{biblio}
\end{document}